\documentstyle [12pt,epsf] {article}

\begin{document}
\begin{titlepage}

\begin{center}
\vspace{2cm}
\LARGE
The Formation of Bulges and Black Holes:\\
lessons from a census of active galaxies in the SDSS\\

\vspace{1cm} 
\large
Guinevere Kauffmann$^1$ and Timothy M. Heckman$^2$\\   
\vspace{0.3cm}
\small
{\em $^1$Max-Planck Institut f\"{u}r Astrophysik, D-85748 Garching, Germany} \\
{\em $^2$Department of Physics and Astronomy, Johns Hopkins University, Baltimore, MD 21218}\\
\vspace{0.6cm}
\end{center}
\normalsize
\begin {abstract}
We examine the relationship between galaxies, supermassive black
holes and AGN using a sample of 23,000 narrow-emission-line (``Type 2'')
AGN drawn from a sample of 123,000 galaxies from the Sloan
Digital Sky Survey. We have studied how AGN host properties
compare to those of normal galaxies and how they depend on the
luminosity of the active nucleus. We find that AGN reside in
in massive galaxies and have distributions of  sizes and
concentrations that are similar to those of the early-type 
galaxies in our sample. The host galaxies of low-luminosity AGN 
have stellar populations similar to normal early-types.
The hosts of high-luminosity AGN have much younger mean
stellar ages and a significant fraction have experienced
recent starbursts. High-luminosity AGN are  also found
in lower density environments. We then use the 
stellar velocity dispersions of the AGN hosts to estimate black
hole masses and their [OIII]$\lambda$5007 emission line       
luminosities to estimate black hole accretion rates.           
We  find that the volume averaged ratio of star formation to
black hole accretion is $\sim 1000$ for the bulge-dominated 
galaxies in our sample. This is remarkably similar to the
observed ratio of stellar mass to black hole mass in nearby
bulges. Most of the present-day  black hole growth  is occurring
in black holes with masses less than  $3 \times 10^7 M_{\odot}$.
Our estimated accretion rates imply that low mass black holes  
are growing on a timescale  that is comparable to the age of the 
Universe Around 50\% this growth takes place in  AGN that are                  
radiating within a factor of five of the Eddington luminosity.  
Such systems are rare, making up only 0.2 \% of the low mass
black hole population at the present day. The rest of growth 
occurs in lower luminosity AGN. The growth timescale 
increases by more than an  order of magnitude for the most 
massive black holes in our sample. We conclude that the    
evolution of the AGN luminosity function documented in recent optical
and X-ray surveys is driven by a decrease in the characteristic mass
scale of actively accreting black holes.

\end {abstract}
\end {titlepage}

\section {Introduction}

Over the past few years there have been remarkable developments
in our understanding of active galactic
nuclei (AGN) and their role in galaxy formation and evolution. 
There is now compelling evidence (Herrnstein et al. 1999; Genzel et al. 2000)
for the existence  of the supermassive black holes
that were long hypothesized
as the power-source for active  galactic nuclei (Salpeter 1964; Lynden-Bell 1969). The local mass
density in these black holes is sufficient to have powered
the known AGN population  
if plausible values for the radiative efficiency of the accreting material are assumed
(Yu \& Tremaine 2002). 
The tight  correlation between the mass of the black hole and the
velocity dispersion and mass of the galactic bulge
within which it resides 
(Ferrarese \& Merritt 2000; Gebhardt et al. 2000; Marconi \& Hunt 2003)
is compelling evidence for a strong connection between
the formation of the black hole and its host galaxy
(e.g. Cattaneo et al. 1999; Kauffmann \& Haehnelt 2000; Granato et al. 2001). Finally, deep
surveys with a generation of powerful X-ray observatories
have established that the  AGN population exhibits so-called ``cosmic down-sizing'':
the space density of  AGN with low x-ray luminosities  peaks at {\em lower redshift} than that
of AGN with high x-ray luminosities
(Steffen et al 2003; Ueda et al. 2003).
These results indicate that a substantial amount of the total black hole growth has occured
more recently than would have been deduced based on 
optical surveys of powerful quasars 
(Boyle et al. 2000; Fan et al. 2001). 

The co-evolution of galaxies and black holes can be directly investigated
through deep optical or x-ray surveys spanning a broad range in redshift.
In our work, we  have taken a complementary approach and we have used the
Sloan Digital Sky Survey (SDSS) to examine the relationship between galaxies,
supermassive black holes and AGN  in the present-day Universe.
Most of the results presented in this review have been published
in a series of papers by  Kauffmann et al. (2003a,b,c;2004) and by Heckman et
al (2004).

\section {Our SDSS Sample}

Up to now, studies  of AGN host galaxies have been limited
by small sample size. In order to carry out detailed  statistical analyses of host galaxy properties,
one requires complete  magnitude-limited samples of galaxies 
with spectra of high enough quality to identify AGN based on the       
characteristics of their emission lines. 

The Sloan Digital Sky Survey
(York et al. 2000; Stoughton et al. 2002)
is using a  dedicated 2.5-meter wide-field
telescope at the Apache Point Observatory to conduct an imaging and
spectroscopic survey of about a quarter of the extragalactic sky. The imaging is
conducted in the $u$, $g$, $r$, $i$, and $z$ bands (Fukugita et al. 1996; Gunn et al. 1998;
Hogg et al. 2001; Smith et. al. 2002),
and spectra are obtained with a pair of multi-fiber spectrographs.
When the current survey is complete, spectra will have
been obtained for nearly 600,000 galaxies and 100,000  QSOs selected uniformly
from the imaging data. Details on the spectroscopic target selection
for the ``main'' galaxy sample and QSO sample can be found in
Strauss et al. (2002) and
Richards et al. (2002). Details about the tiling algorithm
and the astrometry can be found in Blanton et al (2003a) and Pier et
al (2003), respectively.  The results shown here  are   
based on spectra of $\sim$122,000 galaxies with 
$14.5 < r < 17.77$  contained in the the SDSS Data
Release One (DR1; Abazaijan et al 2003). These data were
made publicly available in 2003.

The spectra are obtained
through 3 arcsec diameter fibers. At the median redshift of the main
galaxy sample ($z \sim$ 0.1), the projected aperture diameter  is
5.5 kpc and  typically contains 20 to 40\% of
the total galaxy light. The SDSS spectra are thus closer to global
than to nuclear spectra. At the median redshift the spectra
cover the rest-frame wavelength range from $\sim$3500 to 8500 \AA\
with a spectral resolution $R \sim$ 2000 ($\sigma$$_{instr} \sim$
65 km/s). They are spectrophotometrically calibrated through
observations of F stars in each 3-degree field (Tremonti et al, in preparation).
By design, the spectra are well-suited to the determination
of the principal properties of the stars and ionized gas in galaxies.
The absorption line indicators (primarily the 4000 \AA \hspace{0.1cm} break strength and
the H$\delta_A$ index) and the emission line fluxes analyzed here  are calculated using a 
special-purpose code described in detail in Tremonti et al.

The rich stellar absorption-line spectrum of a typical SDSS galaxy provides 
unique information about its stellar content and  dynamics. 
However, it makes the measurement
of weak nebular emission-lines quite difficult. To deal with this,
we have performed a careful subtraction of the stellar absorption-line
spectrum before measuring the nebular emission-lines. This is accomplished
by fitting the emission-line-free regions of the spectrum with a model
galaxy spectrum computed using   the new population synthesis
code of Bruzual \& Charlot (2003, hereafter BC2003), which incorporates
a high resolution (3 \AA\ FWHM) stellar library. 

We have used the amplitude of the 4000 \AA\ break (the narrow version of the index defined in
Balogh et al. 1999)
and the strength of the H$\delta$ absorption line (the Lick
$H\delta_A$ index of Worthey \& Ottaviani 1997) as diagnostics of the
stellar populations of the host galaxies. Both indices
are corrected for the observed contributions of the emission-lines
in their bandpasses.
Using a library of
32,000 model star formation histories,
we have used the measured $D_n(4000)$ and $H\delta_A$ indices
to obtain a maximum likelihood  estimate of  the  $z$-band mass-to-light ratio for each galaxy.
By comparing the colour predicted by the  best-fit model to the observed colour of the galaxy,
we also estimate the attenuation of the starlight due to dust.
The reader is referred to Kauffmann et al (2003a) for more details.

The SDSS imaging data provide the basic structural parameters that are used in this analysis.
We use the $z$-band as our fiducial filter because it is the least sensitive to the
effects of dust attenutaion.
The $z$-band absolute magnitude, combined with our estimated values of M/L and dust attenuation
$A_z$ yield the stellar mass ($M_*$). The half-light radius in the $z$-band and the
stellar mass yield the effective stellar surface mass-density
($\mu_* = M_*/2\pi r_{50,z}^2$). As a proxy for Hubble type we use
the SDSS ``concentration'' parameter $C$, which is defined as the ratio
of the radii enclosing 90\% and 50\% of the galaxy light in the $r$ band
(see Stoughton et al. 2002). Strateva et al. (2001) find that galaxies
with $C >$ 2.6 are mostly early-type galaxies, whereas spirals and irregulars
have 2.0 $< C <$ 2.6.

\section {Bimodality in the Physical Properties of Galaxies}

Before we discuss the properties of the galaxies that host AGN, it is useful
to review the observed  relations between ''normal'' galaxies in the local Universe.
It is now more than 75 years since Hubble introduced his now-standard classification
scheme for galaxies (Hubble 1926). Hubble realized very early on that galaxies exhibit 
striking regularity in their properties. In the simplest form, Hubble's
scheme recognizes three basic galaxy types: ellipticals, spirals and
irregulars. These can be arranged in a linear sequence along which many
properties vary coherently.

The advent of large redshift surveys of galaxies has  enabled us not only
to study the correlations between different galaxy properties, but also
to quantify the {\em distribution} of galaxies in a multi-dimensional space
of physical parameters. It has become clear that the galaxy population is
strongly bimodal. This is seen in the distribution of galaxies  as
a function of colour; Strateva et al (2001) and Blanton et al (2003b) show
that the colour distribution of galaxies  has two pronounced peaks. 
The clustering properties of galaxies that fall into these two peaks are 
also very different. Red galaxies are  more strongly clustered than
blue galaxies and their two point correlation function is significantly steeper
 on small scales (Zehavi et al 2002; Madgwick et al 2003).

In Kauffmann et al (2003b), we studied the relations between the
stellar masses, sizes, internal structure, and star formation histories of galaxies.
We showed that the galaxy population as a whole divides into two distinct ``families''.
Below a characteristic stellar mass of $\sim 3 \times 10^{10} M_{\odot}$, galaxies
have low surface densities and concentrations typical of disk systems. 
The median surface mass density of  low mass galaxies
scales with mass as $\mu_* \propto M_*^{0.63}$. At larger masses,
the scaling of $\mu_*$ with $M_*$ becomes weaker and $\mu_*$ eventually saturates at
at a value of around $10^9 M_{\odot}$ kpc$^{-2}$ (Fig. 1).  

We also presented the conditional density distributions of two stellar age
indicators, the 4000 \AA\ break strength D$_n$(4000) and the H$\delta_A$
index as functions of stellar mass, stellar surface density and concentration. 
Faint low mass galaxies
with low concentrations and surface mass
densities have young stellar populations, ongoing star formation and blue colours. 
Bright galaxies with high stellar
masses, high concentrations and high surface densities have old stellar populations,
little ongoing star formation and red colours. A transition from ``young'' to ``old'' 
takes place at a characteristic stellar mass of $3\times 10^{10} M_{\odot}$, a 
stellar surface density of $3 \times 10^{8} M_{\odot}$kpc$^{-2}$ and a concentration index
of 2.6 (Fig. 2).

\begin{figure}
\centerline{
\epsfxsize=7cm \epsfysize=7cm \epsfbox{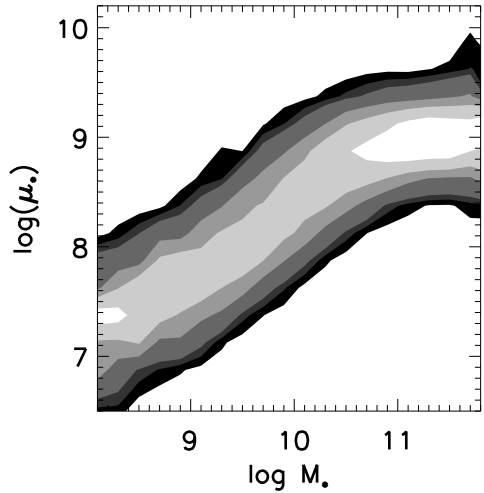}
}
\caption{\label{fig1}
\small
Conditional density distributions showing trends in the structural parameter 
$\mu_*$ 
as a function the logarithm of stellar mass.
Galaxies have been weighted by $1/V_{max}$ and the bivariate
distribution function has been normalized to a fixed number of
galaxies in each bin of $\log M_*$.}    
\end {figure}
\normalsize

\begin{figure}
\centerline{
\epsfxsize=12cm \epsfysize=12cm \epsfbox{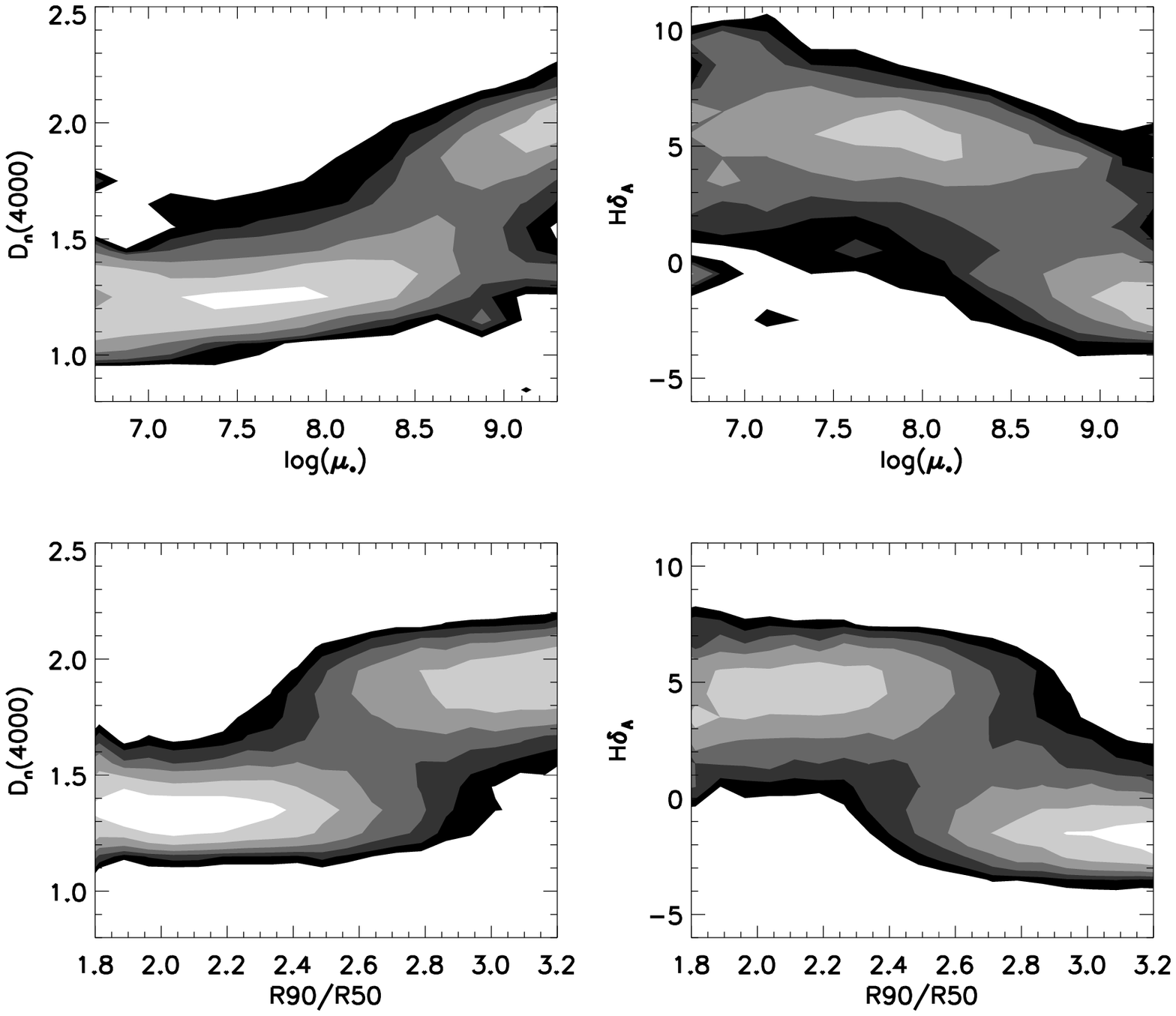}
}
\caption{\label{fig2}
\small
Conditional density distributions showing trends in the stellar age indicators
D$_n$(4000) and H$\delta_A$ 
as functions of the logarithm of the surface mass density $\mu_*$ and 
of the concentration index $C$.}
\end {figure}
\normalsize

\section {Identifying and Classifying AGN in the SDSS}

According to the standard ``unified'' model
(e.g. Antonucci 1993), AGN can be broadly classified into two categories
depending on whether the central black hole and its associated continuum
and broad emission-line region is
viewed directly (a ``type 1'' AGN) or is obscured by a dusty circumnuclear
medium (a ``type 2'' AGN). 
In type 1 AGN the optical continuum is dominated
by non-thermal emission, making it  a challenge to study the host galaxy and its
stellar population.
We have therefore excluded the type 1 AGN from our initial sample.

Baldwin, Phillips \& Terlevich (1981, hereafter BPT) demonstrated that it was possible to distinguish
type 2 AGNs from normal star-forming galaxies by considering the intensity ratios of two pairs
of relatively strong emission lines, and this technique was refined by Veilleux \& Osterbrock
(1987). It has become standard practice to classify objects according to
their position on the so-called BPT diagrams. Fig. 3 shows an example of such a diagram
for all the emission-line galaxies in our sample. We have plotted the ratio
[OIII]$\lambda$5007/H$\beta$ versus the ratio [NII]$\lambda$6583/H$\alpha$ for all galaxies
where all
four lines were detected with $S/N>3$. 

Fig. 3 shows that there are two well-separated sequences of
emission line galaxies.
Based on these data, we have chosen to define the demarcation  between starburst
galaxies and AGN  as follows: A galaxy is defined to be an AGN if
\begin{equation} \log ([\rm{OIII}]/H\beta) >  0.61/(\log ([\rm{NII}]/H\alpha)-0.05) +1.3 \end{equation}    
This curve is represented by a dashed line in Fig.3.

\begin{figure}
\centerline{
\epsfxsize=10cm \epsfbox{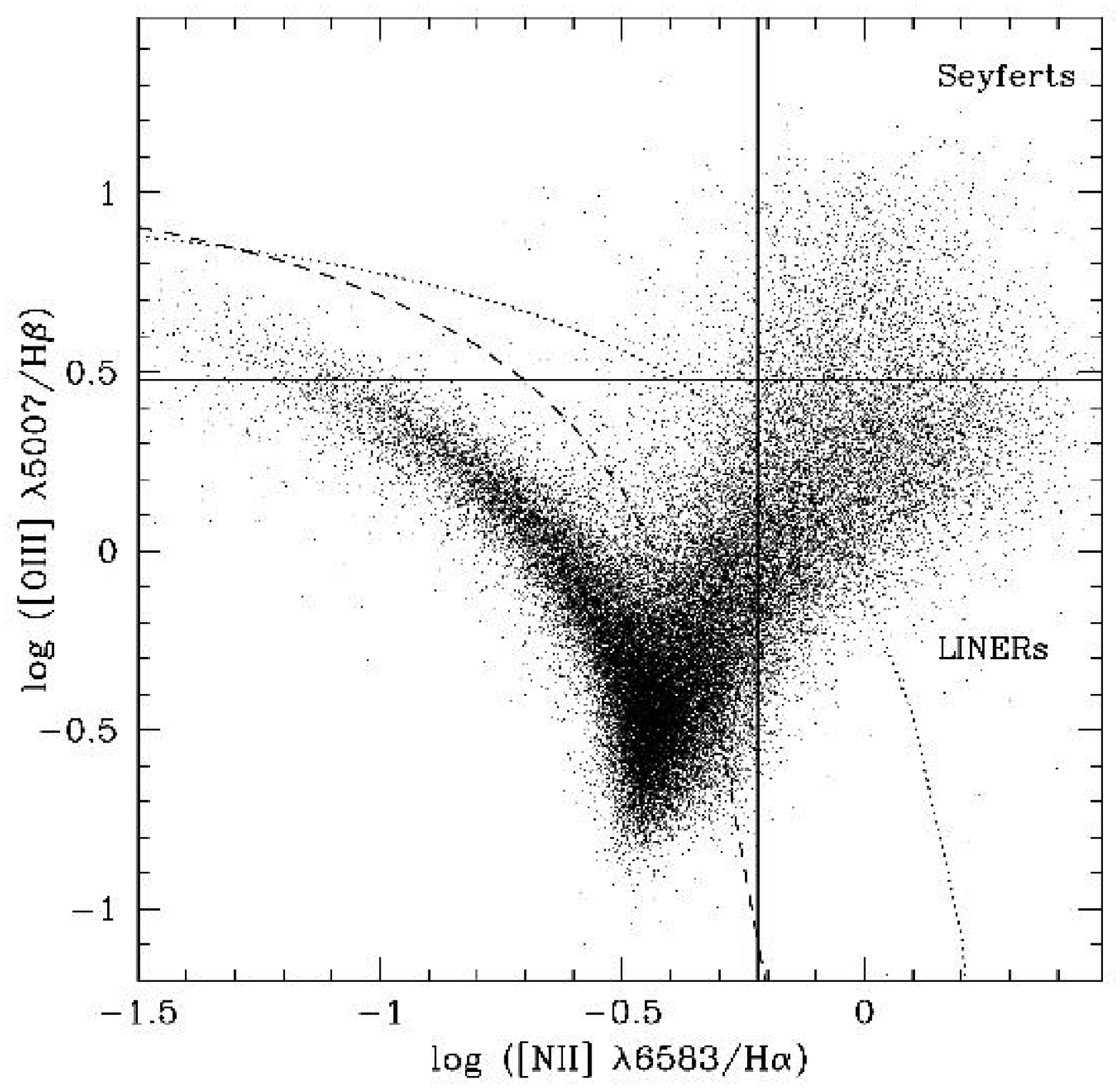}
}
\caption{\label{fig3}
\small
An example of a BPT (Baldwin, Phillips \& Terlevich 1981) diagram in which we plot
the emission line flux ratio [OIII]/H$\beta$ versus the ratio [NII]/H$\alpha$ for all
the galaxies in our sample where all four lines are detected with $S/N >3$ (55,757 objects).
The dotted curve shows the demarcation between starburst galaxies and AGN
defined by Kewley et al (2001). The dashed curve shows our revised demarcation (equation 1). 
A total of 22,623 galaxies lie above dashed curve.
Seyfert galaxies are often defined to have [OIII]/H$\beta > 3$ and 
[NII]/H$\alpha>0.6$, and LINERs to have [OIII]/H$\beta < 3$ and [NII]/H$\alpha > 0.6$.
Our sample includes 2537 Seyferts and 10,489 LINERs according to this definition.}
\end {figure}
\normalsize

Narrow-line (type 2)
AGN are traditionally divided into  3 general classes:
type 2 Seyferts, LINERs, and the so-called ``transition'' objects. 
In traditional classification schemes (see for example Ho, Filippenko \& Sargent 1997), Seyfert galaxies
are identified as those objects with high values of both [OIII]/H$\beta$ ($>3$) and of other ratios 
involving lower ionization lines, such as [NII]/H$\alpha$, [SII]/H$\alpha$ and [OI]/H$\alpha$.
LINERs, on the other hand, have lower values of [OIII]/H$\beta$ ($< 3$), but high values of
ratios involving the lower ionization lines.

A contribution to the emission-line spectrum by
both star-formation and an AGN is almost inevitable in many of the
SDSS galaxies, given the relatively large projected aperture size
of the fibers (5.5 kpc diameter at $z=0.1$).   
This is much larger than the $\sim$200 pc apertures used in the
survey of nearby galaxy nuclei by Ho, Filippenko \& Sargent (1997). It is
therefore not
surprising that the majority of AGN in our sample fall into the ``transition''
class, and have line ratios intermediate between those of
star-forming galaxies and those of LINERs or  Seyferts.

We therefore prefer an AGN classification system that is less sensitive to
aperture, while still reflecting the difference in intrinsic nuclear luminosity
between Seyfert galaxies and LINERs. 
In our analysis, we focus on the luminosity of
the [OIII]$\lambda$5007 line as a tracer of AGN activity. (Note
that we correct the [OIII] luminosity for extinction using
the Balmer decrement.)  Although this line can be excited by
massive stars as well as an AGN,
it is  known to be relatively 
weak in metal-rich, star-forming galaxies.   
The virtue of classifying galaxies by [OIII] luminosity is that this then
allows us to study the large 
number of transition galaxies with spectra intermediate between pure 
star-forming systems and pure LINERs/Seyferts. 
We  stress that any attempt to characterize the stellar
population in  AGN hosts  must  include
the transition objects that comprise the majority of the AGN in
Fig. 3. Excluding these would bias the sample against host galaxies
with significant amounts of on-going star-formation.

\section {The Masses and Structural Properties of AGN Host Galaxies}

\subsection {AGN are predominantly in massive galaxies} 

The AGN in our sample are drawn from an $r$-band limited spectroscopic survey of galaxies.
Galaxies of different luminosities can be seen to different distances
before dropping out due to the selection limits of the survey.
The volume $V_{max}$  within which a galaxy can be seen and will be included in the sample
goes as the distance limit cubed, which results in the     
samples being dominated by intrinsically bright galaxies.
In our work, we correct for this effect by giving each galaxy or AGN
a weight equal to the inverse of its maximum
visibility volume determined from the apparent magnitude limit of the survey.

Our sample of emission-line galaxies includes 55,747 objects where
the four emission lines, H$\alpha$, [NII], [OIII] and H$\beta$,
are all detected with $S/N >3$. In total, forty percent of these galaxies
are classified as AGN, but this number is strongly dependent on host mass and to
a lesser extent, on morphological type.
As can be seen, AGN are preferentially found in more massive and more
concentrated galaxies. The dependence  of AGN fraction on mass
is very  striking. Even though 70-80\% of galaxies
with stellar masses less than $10^{10} M_{\odot}$ have detectable emission lines,
only a few percent are classified as AGN. In contrast, more than
80\% of emission line galaxies with $M_* > 10^{11} M_{\odot}$
are AGN.  

It should be noted that AGN detection rates are subject to strong selection effects.
Ho et al (1997) find that
43\% of the galaxies in their survey can be regarded as active and nearly all
their galaxies have detectable emission lines.
In our sample, the nuclear spectrum will be  
diluted by the light from  surrounding host galaxy and our derived AGN fractions in
massive galaxies  are a
strong function of distance. For nearby galaxies, the AGN fraction  reaches 
50\%, a value very similar to the fraction that  Ho et al found for
the $L_*$ galaxies in their sample.
On the other hand, the AGN fraction in low mass galaxies does not rise above
a few percent, even for nearby galaxies. There are  6586 galaxies in our sample
with stellar masses in the range $10^8-3\times 10^9 M_{\odot}$, so this result has high
statistical significance.  

We find that for [OIII] luminosities greater than $10^7$ $L_{\odot}$ , 
the AGN fraction no longer depends on distance.
We will henceforth refer to AGN as ``strong'' or ``weak'' using
L[OIII] = $10^7 L_{\odot}$ as a dividing line. The fraction of strong AGN in galaxies 
with masses in the range  $3 \times 10^{10} -10^{11} M_{\odot}$ is $\sim 0.1$.
There are fewer strong AGN in both lower mass and higher mass galaxies (see
Kauffmann et al (2003c)).

\subsection {Structural properties resemble those of  normal early-type galaxies}

Most previous studies of the global morphologies 
of type 2 AGN hosts have concentrated on  
Hubble type.     
It should be noted that the standard morphological
classification scheme defined by Hubble  mixes elements that depend on the structure of a
galaxy (disk-to-bulge ratio, concentration, surface density) with
elements related to its recent star formation history (dust-lanes,
spiral arm strength). 
Because it is impractical to classify hundreds of thousands of galaxies by eye,
studies of galaxy morphology  in the SDSS have focused on                       
simple structural parameters that can be measured automatically for
very large numbers of objects. In section 2, we defined two  such parameters: 
the concentration index $C$ and the stellar surface mass density $\mu_*$.

In Fig. 4, we compare the surface mass density distributions of AGN hosts to those
of ordinary early-type and late-type galaxies.
We show results for four different ranges in $\log M_*$.
We have separated early-type galaxies from late-types  at a $C$ index of 2.6.
As shown in Fig. 2, this is where there is a pronounced ''jump''
from a population of galaxies with predominantly  young stellar populations to a
population with mainly old stars.    
Fig. 4 shows that the surface density distributions of AGN host galaxies strongly resemble
those of {\em early-type galaxies} of the same stellar mass.

\begin{figure}
\centerline{
\epsfxsize=12.5cm \epsfbox{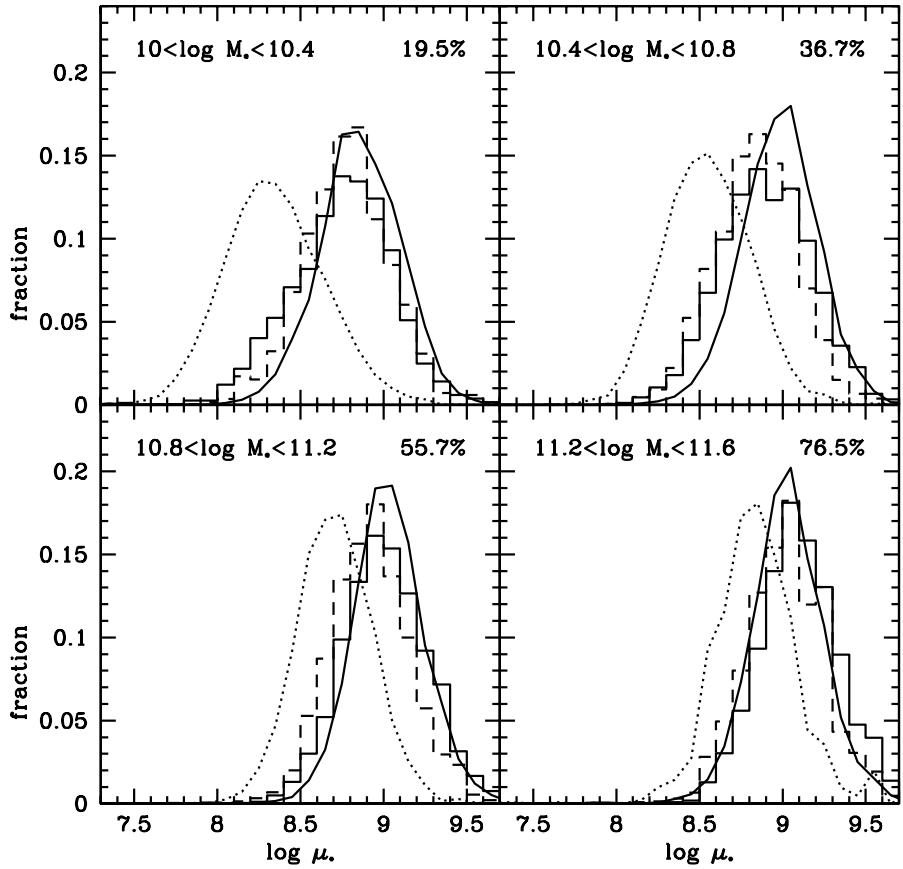}
}
\caption{\label{fig4}
\small
The surface mass density distributions of the host galaxies of weak AGN (solid black
histogram) and
of strong AGN (dashed black histogram) are compared to
those of early-type ($C>2.6$) galaxies [solid curve] 
and late-type ($C<2.6$) galaxies [dotted curve]
in 4 separate ranges of $\log M_*$. The numbers listed in the top
right corner of each panel indicate the percentage of galaxies   
in that mass range that are early-type.}
\end {figure}
\normalsize

\section {The AGN-Starburst Connection }

\subsection {Diagnostics of Young Stars}

For young massive stars, the strongest spectral features with the
greatest diagnostic power lie in the vacuum ultraviolet regime
between the Lyman break and $\sim$2000 \AA\ (e.g.
Leitherer et al. 1999; de Mello et al. 2000). These include the
strong stellar wind lines of the OVI, NV, SiIV, and CIV resonance
transitions and a host of weaker stellar photospheric lines.
Most of the photospheric lines arise from highly-excited states
and their stellar origin is unambiguous. While resonance absorption lines
may have an interstellar origin, the characteristic
widths of the stellar wind profiles make them
robust indicators of the presence of massive stars. Unfortunately,
observations in this spectral regime are difficult. Only a handful
of local type 2 Seyferts and LINERs have nuclear UV fluxes that are 
high enough to enable
a spectroscopic investigation. While this small sample may not be
representative, the available data firmly establish the presence
of a dominant population of young stars (Heckman et al. 1997;
Gonz\'alez Delgado et al. 1998; Maoz et al.1998; Colina \& Arribas 1999).

While the optical spectral window is far more
accessible, the available diagnostic features
of massive stars are weaker and less easy to interpret.
Old stars
are cool and have many strong spectral features in the optical
due to molecules and low-ionization metallic species.
Hot young stars have relatively featureless
optical spectra.
Thus, the spectroscopic impact of the presence of young stars is mostly
an indirect
one: as they contribute an increasing fraction
of the light (as the luminosity-weighted mean stellar age
decreases) most of the strongest spectral features in the optical weaken. 
This effect is easy to measure. 
Unfortunately, the effect of adding ``featureless'' nonstellar continuum
from an AGN and young starlight will be similar in this regard.

The strongest optical absorption lines from young stars are
the Balmer lines. These reach peak strength in early A-type stars,
and so they are most sensitive to a stellar population with an age of
$\sim$ 100
Myr to 1 Gyr (e.g. Gonz\'alez Delgado et al. 1998).
Thus, the Balmer lines do not uniquely
trace the youngest stellar population. 
On the plus side, they can
be used to characterize past bursts of
star formation (e.g Dressler \& Gunn 1983; Kauffmann et al 2003a).

In the near-IR, red supergiants will contribute significantly
to the light from young stellar populations.
The spectral features produced by red supergiants are qualitatively
similar to those produced by red giants that dominate the near-IR
light in an old stellar population. A robust method
to determine whether old giants or young supergiants dominate
is to measure the M/L ratio in the near-IR using the stellar velocity
dispersion. So far this technique has been applied to only a small
sample, but the results are tantalizing (Oliva et al. 1999;
Schinnerer et al. 2003). 
Another way of diagnosing the presence of young stars is to make
use of spectral information in both the optical and the near-IR.
In an old stellar population, cool stars dominate                      
and so the associated metallic and
molecular spectral features are strong in both bands. In contrast,
the optical (near-IR) continuum in a young stellar population will
be dominated by hot main sequence stars (cool supergiants). The
metallic/molecular lines are therefore weak in the optical and strong
in the near-IR. This combination of properties provided
some of the first direct evidence for a young stellar population in AGN
(Terlevich et al. 1990; Nelson \& Whittle 1999).

\subsection{Young Stars in Active Galactic Nuclei: background}

Spectroscopy of the nearest AGN affords the opportunity to study
the starburst-AGN connection on small physical scales
($>$ a few pc).
The drawback is that the nearest AGN have low-luminosities, and we might
expect that the amount of star formation associated with black-hole
fueling would scale in some way with AGN power. To get a complete picture
it is thus important to examine both the nearest AGN and more 
powerful AGN. 

The earliest investigation of the stellar population for a moderately
large sample of the nearest AGN was by Heckman (1980a,b), who discussed
30 LINERs
found in a survey of a sample of 90
optically-bright galaxies.
The typical projected radius of the spectroscopic aperture was $\sim$200 pc.
The spectra covered the range from 3500 to 5300 \AA. 
LINERs were primarily found in galaxies of early
Hubble type (E through Sb). Based on the strengths of the
stellar metallic lines and the Balmer lines, the
nuclear continuum was dominated by old stars in about 3/4 of the LINERs,
while a contribution of younger stars was clearly present in the remainder.
Typical luminosities of the [OIII]$\lambda$5007 and H$\alpha$
NLR emission lines were
$\sim 10^5$ to $10^6 L_{\odot}$.

Ho, Filippenko, \& Sargent (2003)
recently examined the nuclear (typical radius
$\sim10^2$ pc) stellar population in a complete sample of $\sim$500
bright, nearby galaxies (of which 43\% contain an AGN).
The larger sample size
and improved treatment of the emission lines allowed Ho et al. to
study statistically-significant samples of low-luminosity LINERs,
type 2 Seyferts, and Transition nuclei and to span a larger range
in AGN luminosity ($L_{H\alpha} \sim 10^4$ to $10^7 L_{\odot}$).
The disadvantage
of these spectra is that they did not extend shortward
of 4230 \AA\, and so they missed the H$\delta$ and
higher-order Balmer lines that most effectively probe young
stars. Nevertheless, their results
are qualitatively consistent with Heckman (1980a,b). 
Cid Fernandes
et al. (2004) have analyzed the near-UV spectra of 51 low luminosity AGN
from the Ho, Filippenko, \& Sargent (2003) sample. The found that
strong Balmer absorption lines are present in about half of the
Transition nuclei but in only about 10\% of the LINERs. 
A similar result was obtained by Gonz\'alez Delgado et al. (2004)
for 28 AGN from the same sample, but using HST to probe the stellar
population within radii of tens of pc from the black hole.
These results are
consistent with the idea that in the Transition nuclei
both young stars and an AGN contribute to the ionization
of the emission-line gas.

It was recognized very early-on
(e.g. Koski 1978) that the optical spectra of powerful type 2 Seyfert nuclei 
could not be explained purely by an old stellar population. An additional
``featureless continuum'' that typically produced
10\% to 50\% of the optical continuum
was present. Until relatively recently, it was tacitly assumed that
this component was light from the AGN (plausibly light from
the hidden type 1 Seyfert nucleus that had been reflected into our
line-of-sight by free electrons and/or dust). However, a detailed
spectropolarimetric investigation by Tran (1995) and related arguments
by Cid Fernandes \& Terlevich (1995) and Heckman et al. (1995)
showed that
young stars were the majority contributors.
These results have been confirmed by several major optical spectroscopic investigations
of moderately large samples of type 2 Seyfert nuclei
(Schmitt, Storchi-Bergmann, \& Cid Fernandes 1999;
Gonz\'alez Delgado, Heckman, \& Leitherer 2001; Cid Fernandes et al. 2001;
Joguet et al. 2001). 
The principal conclusion is that a young 
($<$ Gyr) stellar population is clearly present in about half
of the Seyfert 2 nuclei. Cid Fernandes et al. (2001) found that
the fraction of nuclei with young stars is 
$\sim$60\% when $L_{[OIII]} > 10^7 L_{\odot}$.
They also found that the ``young'' Seyfert 2 nuclei were
hosted by galaxies with much larger far-IR luminosities than the ``old'' nuclei,
suggesting that the global star formation rate was higher.
\subsection{Results from SDSS}

One of the great advantages of studying the properties of 
AGN in the SDSS is that one can compare the results for AGN hosts
with those derived for ``normal'' galaxies in a very precise way.
In Fig. 4, we showed that AGN 
occupy host galaxies
with structural properties similar to ordinary early-type galaxies.
The properties of the stellar populations of normal 
galaxies are known to correlate  
strongly with morphological type. Early-type galaxies have old stellar
populations and very little gas and dust, whereas there is usually plenty of   
ongoing star formation in late-type galaxies. 
In Fig. 5, we compare the D$_n$(4000) distributions of AGN hosts with
`normal' early-type ($C>2.6$) and late-type ($C<2.6$) galaxies. This plot 
reaches a rather different conclusion to Fig. 4:
the D$_n$(4000)  distributions of  weak AGN  resemble those of 
early-type galaxies,         
but that strong AGN have stellar ages similar to those of late-type galaxies.               
One interpretation of these results 
is that powerful  AGN reside in early-type galaxies that are
undergoing or have recently undergone a
transient star-forming event (starburst). The magnitude of this event
might plausibly scale with AGN luminosity, so that its effects on
the stellar population and structural parameters are most pronounced
in the hosts of strong AGN. 
If this hypothesis is correct, then we ought to see
evidence that the
star formation timescales in high luminosity AGN are short.

Strong H$\delta$ absorption arises in galaxies that 
experienced a burst of star formation that ended
$0.1-1$ Gyr ago. As discussed in Kauffmann et al (2003a), 
the location of galaxies in the
D$_n$(4000)/H$\delta_A$ plane is a powerful diagnostic 
of whether they have been forming
stars continuously or in bursts over the past 1-2 Gyr. Galaxies with continuous star formation
histories occupy a very narrow strip in this plane. 
 A recent burst is {\em required} in order
to obtain  significant displacement away from this locus. 
In Fig. 6
we plot the fraction $F$  of AGN with H$\delta_A$ values that are displaced
by more than 3$\sigma$ above the locus of continuous models as a a function
of [OIII] luminosity.
This fraction increases from several percent for the weakest AGN to nearly
a quarter for the strongest.
The corresponding value of $F$ for the sample of     
normal massive galaxies is only 0.07. For normal massive galaxies with
`young' stellar populations (D$_n$(4000)$<1.6$), $F=0.12$.

\begin{figure}
\centerline{
\epsfxsize=8.5cm \epsfbox{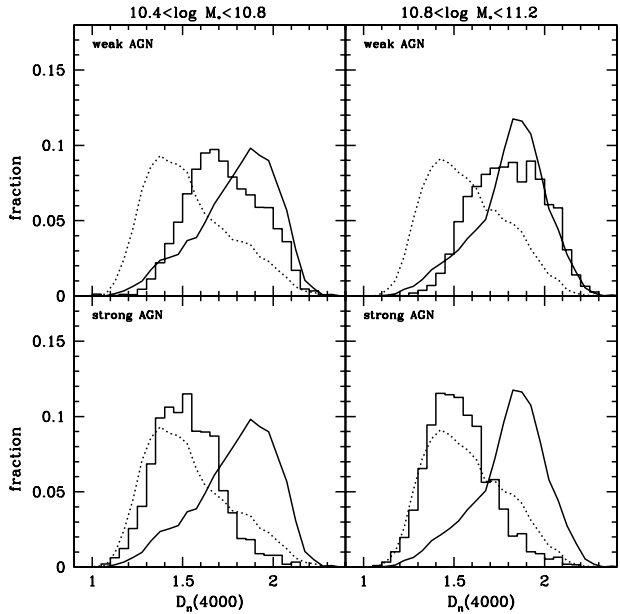}
}
\caption{\label{fig5}
\small
Top: The D$_n$(4000) distribution for the  host galaxies
of weak AGN with log L[OIII]$ < 7.0$  (histogram) is compared to
those for  early-type ($C>2.6$) galaxies (solid curve) and late-type 
($C<2.6$) galaxies (dotted curve)
in 2 separate ranges of $\log M_*$. Bottom: The same, except for
strong AGN with log L[OIII]$ >7.0$}
\end {figure}
\normalsize

\begin{figure}
\centerline{
\epsfxsize=6cm \epsfbox{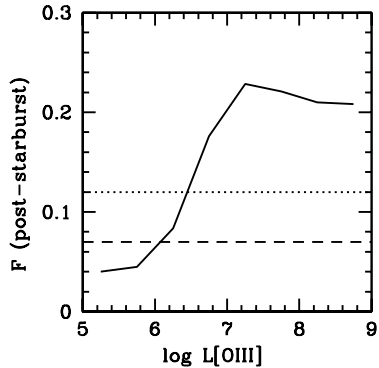}
}
\caption{\label{fig6}
\small
The fraction $F$  of AGN with H$\delta_A$ values that are displaced
by more than 3$\sigma$ above the locus of star-forming galaxies is
plotted as a function of log L[OIII].
The dashed line indicates the fraction of such systems in the
subsample of normal massive galaxies. The dotted line
indicates the fraction of such systems in the subsample
of normal massive galaxies with D$_n$(4000)$< 1.6$.}
\end {figure}
\normalsize

As discussed above, most previous studies of the stellar 
populations of type 2 AGN hosts have focused on the
{\em nuclear regions} of the host galaxies.
We obtain qualitatively similar trends in 
mean stellar age as a function of 
AGN luminosity to these previous studies, but
we find that a higher fraction of AGN
with [OIII] luminosities greater than $10^7 L_{\odot}$ contain young
stellar populations.
One might thus speculate that there might be  systematic radial
gradients in the stellar populations of the host galaxies of these systems.
We have  tested this by splitting our sample into different bins in normalized
distance $z/z_{max}$.
We then find that there are
rather strong radial gradients in mean stellar age and that the youngest stars
appear to be located  well {\em outside} the  nuclei of the host galaxies.

\section {Towards a Physical Interpretation}

In the previous section, we showed that there is a strong correlation of 
age-sensitive stellar absorption indices  with AGN luminosity. Only 
AGN with the weakest [OIII] emission  have stellar ages in the range that is normal for 
early-type galaxies (D$_n$(4000)$ > 1.7$ and H$\delta_A <$ 1) (Fig. 7)
Our results are tantalizing in that they show that AGN activity
and star formation are closely linked in galaxies. In this section we attempt
to take our analysis one step further and study the relation between
star formation and accretion onto black holes in a more quantitative 
way.

\begin{figure}
\centerline{
\epsfxsize=10cm \epsfbox{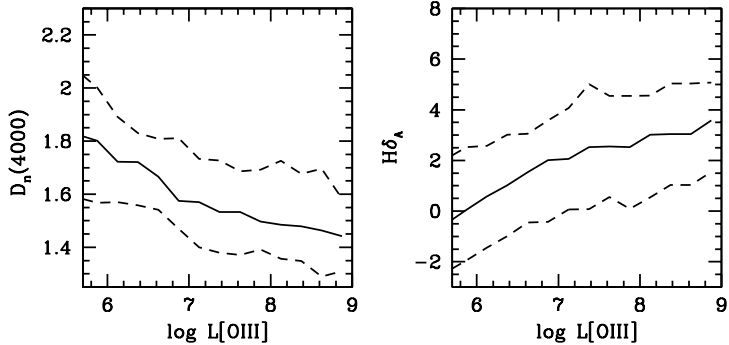}
}
\caption{\label{fig7}
\small
D$_n$(4000) and H$\delta_A$ are  plotted as a function of log L[OIII]. The solid
line shows the median, while the dashed lines indicate the 16-84 percentiles of the
$1/V_{max}$ weighted distribution.}
\end {figure}
\normalsize

In order to do this, we need to be able to transform from {\em observed} quantities
such as 4000 \AA\ break strength, galaxy mass or
velocity dispersion, and [OIII] luminosity  to {\em physical} quantities, such as 
star formation rate, black hole mass and black hole accretion rate.
We have made the following assumptions:
\begin{enumerate}
 \item
We have used the luminosity of the [OIII]$\lambda$5007 emission line to derive an 
estimate of the AGN luminosity, assuming that the bolometric correction for the [OIII] 
luminosity is the same as that determined for Type 1 AGN. 
This approach is motivated 
by the unified model for AGN, and has been empirically validated in the literature.  
(Note that we do not correct the [OIII] line for extinction when we derive the
bolometric correction.   
We will use the notation $L_{O3}$ to differentiate this ``raw''
[OIII] luminosity  from the extinction-corrected
[OIII] luminosity, which we denote  L[OIII].)
We have used the extensive SDSS samples
of type 1 Seyfert nuclei and low-redshift quasars described in Zakamska
et al. (2003) and Kauffmann et al (2003c) to                                 
derive  an average ratio
$L_{5000}/L_{O3} \sim$ 320 (where $L_{5000}$ is the monochromatic
continuum luminosity $\lambda P_{\lambda}$ at 5000 \AA\ rest-frame). 
We obtain a very similar result using a lower redshift sample of
type 1 Seyfert nuclei ($<z> \sim$ 0.03) compiled by Dahari \&
De Robertis (1988). 
The mean type 1
AGN spectral energy distribution in Marconi et al. (2004) then yields
corresponding bolometric corrections $L_{\rm Bol}/L_{O3} \sim$ 3500.
\item 
We have calculated  black hole accretion rates for an assumed radiative efficiency of 
10\%.  
\item
We have used the stellar velocity dispersion measured within the central (typically $\sim$ 
3 to 10 kpc) region to estimate the black hole mass, using the $M_{BH}-\sigma$ 
relation of Tremaine et al. (2002). 
We only derive black hole masses for bulge-dominated galaxies with
$\mu_* > 3 \times 10^8 M_{\odot}$ kpc$^{-2}$.
\item We have used the methodology described in Brinchmann et al (2004)
to derive star formation rates for the galaxies and AGN
in our sample. In  the region sampled by the
spectroscopic fiber,  star formation rates for non-AGN  are estimated using 
extinction-corrected Balmer emission lines. 
Star formation rates for AGN are derived using the  
tight correlation between $D_n(4000)$ and specific
star-formation rate $SFR/M_*$ that is observed  for normal galaxies. The correction from
SFR inside the fibre  to total star formation
rates  is made using   the $g-r$ and $r-i$ colors
of the galaxies in the region  outside the fiber. 
We remind reader that the projected fiber diameter ranges 
from  a few to ten kpc and typically
encloses 20 to 50\% of the light.
\item
We have considered only Type 2 (narrow line) AGN in which the intense radiation 
from the central accretion disk is completely obscured along our line-of-sight. The effect 
of the missing Type 1 AGN is relatively small (factor of $\sim$2) for our overall global 
assessments.  
\end {enumerate}

\subsection{Which Black Holes are Growing?}

Because  powerful AGN are rare
and the timescale over which black holes accrete most of their mass is
likely to be considerably shorter than the timescale
over which the surrounding galaxy forms its stars,
it is more meaningful to work with {\em volume-averaged} quantities than to present results
for individual AGN.

In Fig. 8 , we plot the ratio of the integrated $L_{O3}$  luminosity in AGN  to the integrated black hole mass 
as a function of $\log M_{BH}$.
This figure shows that most of the present-day accretion is occurring
onto black holes  with $M_{BH} <$ few $\times  10^7 M_{\odot}$. Above this mass, the 
integrated $L_{O3}$  per unit $M_{BH}$ drops dramatically. If we convert $L_{O3}$ 
into a mass accretion rate using the relations described above, 
we find that low mass black holes ($ <$ few $ \times  10^7 M_{\odot}$) 
have a growth time of $\sim 20-30 $ Gyr, or around 2-3 times the age of the Universe.
For more massive black holes ($> 10^8 M_{\odot}$), the growth timescales quickly
increase to orders of magnitude larger than the Hubble time. This indicates that
massive black holes must have formed at significantly higher redshifts
are they are currently experiencing very little additional growth.

As well as looking at the integrated accretion onto black holes of given mass,
it is also interesting to study the {\em distribution} of accretion rates in these systems. This is
shown in Fig. 9. 
We normalize the accretion rate by dividing by the Eddington rate for each object. 
The left panel shows the cumulative fraction of black holes that are accreting above a given
rate. Results are shown for different ranges in black hole mass, with mass
increasing from the solid curve on the right ($3 \times 10^6 M_{\odot}$)
to the dotted curve on the left ($10^9 M_{\odot}$) by a factor of 3 in each case. 
Fig. 9 shows that the 
accretion rate functions cut off fairly neatly at 
$\dot{M}_{BH}/\dot{M}_{edding} \sim 1$. This gives us
confidence that our conversions from $L_{O3}$  to 
accretion rate and from $\sigma$ to black hole
mass are yielding reasonable answers.
There are more low mass black holes than high mass black holes with very large 
accretion rates near the Eddington limit. Roughly 0.5 percent of black holes with 
  $M_{BH} < 3 \times 10^7 M_{\odot}$
are accreting above a tenth Eddington of the Eddington limit.  
 For black holes with 
$M_{BH} \sim 3 \times 10^8 M_{\odot}$, the fraction accreting
above a tenth Eddington has dropped to $10^{-4}$.

The right panel shows the cumulative  fraction of the accreted mass 
as a function of $\dot{M}_{BH}/\dot{M}_{edding}$. Results are
shown for low mass black holes with $M_{BH} < 3 \times 10^7 M_{\odot}$ (solid) and for
high mass black holes with $M_{BH} > 3 \times 10^7 M_{\odot}$ (dashed).
As can be seen, most of the black hole growth is occurring in the
most luminous AGN. For low mass black holes, 50 \% of the growth occurs in AGN
that are within a factor of 5 of the Eddington luminosity. For high mass black holes, half  
of the present-day growth occurs in AGN that are radiating above 
$\sim 8$\%  Eddington.
Yu \& Tremaine (2002) have shown that the  mass density in black holes
estimated from integrating the luminosity function of 
quasars over all cosmic epochs agrees rather well
with the total mass density in  black holes in galaxies at the present day.  
Their analysis did not include low-luminosity or type  2 AGN and it has thus been something
of a puzzle why these two estimates of black hole mass density should agree so well. 
Our analysis demonstrates
that
although low-luminosity AGN are  numerous, their contribution to the growth of
black holes is not dominant.                                                              

\begin{figure}
\centerline{
\epsfxsize=8cm \epsfbox{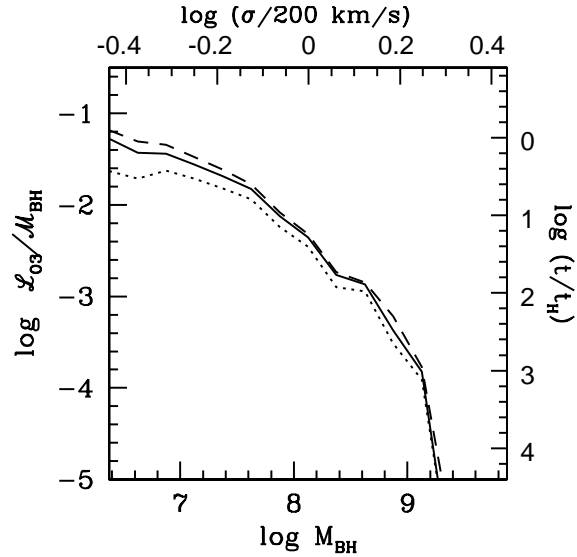}
}
\caption{\label{fig8}
\small
The logarithm of the  ratio of integrated $L_{O3}$  luminosity in AGN
(in solar units) to integrated black hole mass ($M_{BH}$) is plotted as
a function of velocity dispersion $\sigma$ (upper axis)
and  $\log M_{BH}$ (lower axis). The inferred growth time of black
holes in units of the Hubble time is plotted on the right-hand axis.}
\end {figure}
\normalsize

\begin{figure}
\centerline{
\epsfxsize=12cm \epsfbox{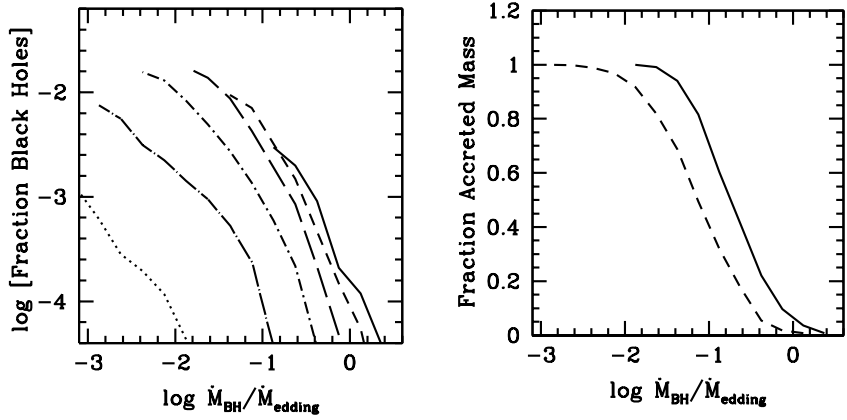}
}
\caption{\label{fig9}
\small
Left: The  fraction of black holes that are accreting above a given rate.
Solid, short-dashed, long-dashed,short dashed-dotted, long dashed-dotted and dotted
curves are for black holes with $3 \times 10^6$, $10^7$, $3 \times 10^7$, $10^8$,
$3 \times 10^8$ and $10^9 M_{\odot}$, respectively.
Right: The fraction of the accreted mass contributed by AGN accreting above a given
rate. The solid curve is for black holes with $M_{BH} < 3 \times 10^7 M_{\odot}$
and the dashed curve is for black holes with masses above this value.}
\end {figure}
\normalsize

\subsection{Where are Black Holes Growing?}

In  Fig. 10, 
we plot the fraction of the integrated $L_{O3}$ luminosity from AGN                           
as a function of stellar mass $M_*$, stellar surface mass density $\mu_*$,
concentration index $C$ and 4000 \AA\ break strength D$_n$(4000).
Fig. 10 shows that most of the present-day accretion is taking place in                     
galaxies with young stellar ages (D$_n$(4000)$< 1.6$),
intermediate stellar masses ($10^{10}$ -few $\times 10^{11} M_{\odot}$),
high surface mass densities ($3 \times 10^8 - 3 \times 10^9$ $M_{\odot}$ kpc$^{-2}$),
and intermediate concentrations ($C \sim 2.6$).
It is quite remarkable that the $L_{O3}$  emission peaks just beyond the
the {\em transition values}  
of $M_*$, $C$ and $\mu_*$ where galaxies appear to undergo a rapid
change from young, star-forming and disk-dominated  to
old, quiescent and bulge-dominated (see Figures 1 and 2).

\begin{figure}
\centerline{
\epsfxsize=8cm \epsfbox{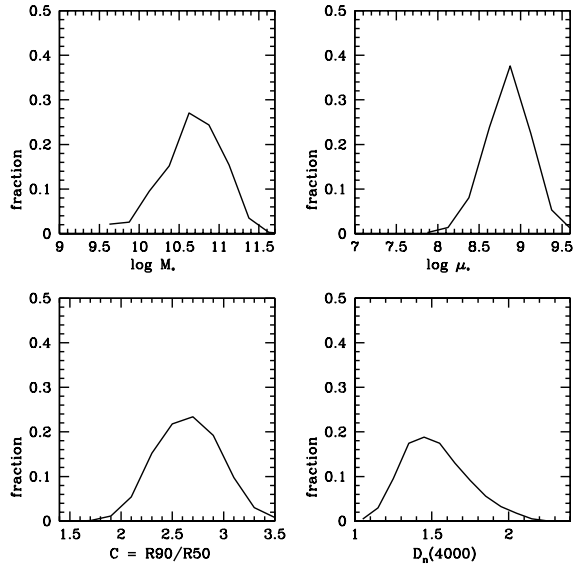}
}
\caption{\label{fig10}
\small
The fraction  of the integrated $L_{O3}$ luminosity in AGN
is plotted as a 
function of $\log M_*$, $\log \mu_*$, $C$, and  D$_n$(4000).}
\end{figure}
\normalsize

\subsection {Bulge building: the relation between star formation and accretion onto black holes}

We have shown that low mass black holes are  currently ``forming'' on a timescale
that is comparable to the age of the Universe.
If the very tight relation
between black hole mass and bulge mass is to be understood, it 
follows that the host bulges of these systems must also be ``forming'' at a comparable rate.

In Fig. 11, we plot  the ratio of the volume-averaged star formation rate 
in galaxies to the volume-averaged accretion
rate onto black holes traced by $L_{O3}$  in AGN. 
The thick black line shows what is obtained if one only considers the SFR 
inside the SDSS fiber aperture.                               
The thin black line shows the result using our
estimates of  {\em total} SFR.  
We have chosen to plot SFR/$\dot{M}_{BH}$ 
as a function
of black hole mass $M_{BH}$   and surface mass density $\mu_*$. From Fig. 11 it is clear
that black hole growth is closely linked to star formation in the bulge.
At low values of  $\mu_*$ characteristic of disk-dominated galaxies,
the ratio of SFR/$\dot{M}_{BH}$ rises
steeply. This is because very few of these galaxies host AGN, but there is plenty of star formation       
taking place in galaxy disks. 
At values of  $\mu_*$ above $3 \times 10^8 M_*$
kpc $^{-2}$ , SFR/$\dot{M}_{BH}$  remains roughly constant. Moreover, it has a value
$\sim 1000$, which is in excellent
agreement with the empirically-derived  relation between black
hole mass and bulge mass (Marconi \& Hunt 2003). 
Given the uncertainties
in the  bolometric correction, in the conversion from $L_{{\rm bol}}$ to
black hole accretion rate and in the estmate of SFR in our AGN, we find it
remarkable that SFR/$\dot{M_{BH}}$  comes out within a factor of a few
of the value that is expected from the $M_{BH}$--$M_{bulge}$ relation.

\begin{figure}
\centerline{
\epsfxsize=11cm \epsfbox{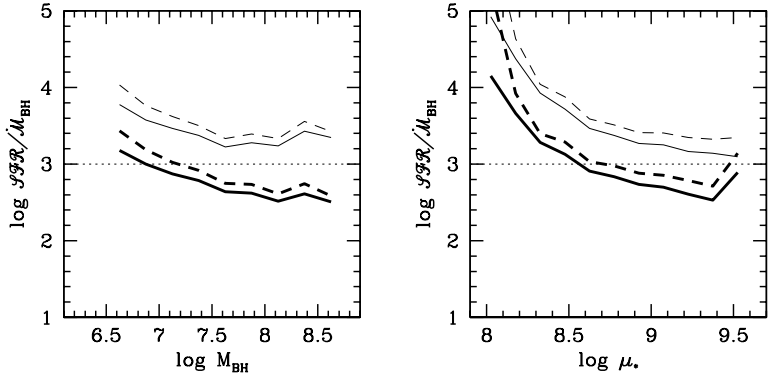}
}
\caption{\label{fig11}
\small
The ratio of the volume-averaged star formation rate in galaxies to the
volume averaged accretion rate onto black holes in AGN is plotted
as a function of $\log M_{BH}$ and $\log \mu_*$. The thick black line shows the
result if SFR is calculated within the fibre aperture for each galaxy
and the thin black lines shows the result using estimates of {\em total} SFR.}
\end {figure}
\normalsize

\section {Discussion}

Our results demonstrate that the volume-averaged  growth of black holes 
through accretion and the volume-averaged growth 
of galaxy bulges through star formation remain coupled at the present time.  
It is therefore likely that the processes that established the 
tight correlation between bulge mass and black hole mass in present-day galaxies
are still operating in low redshift AGN. 

We have also demonstrated that the growth time of low mass black holes ($<3 \times 10^7 M_{\odot}$) derived
from the ratio of the mass density in black holes to the present-day  black 
hole accretion rate is short  (several Hubble times). 
By contrast,  massive black holes have very long growth times. On average, they are 
accreting at a 
rate that is several orders-of-magnitude lower than their past averaged one. This population 
evidently formed early on and has evolved little since then. 

Our results thus lead to a picture in which both star formation and black 
hole growth in galaxies has been steadily moving to lower and lower mass scales since 
a redshift of $\sim$2.  Deep surveys that probe the evolution of the AGN and galaxy 
populations (Cowie et al 1996; Ueda et al. 2003;  Steffen et al 2003) have provided direct 
evidence for this process, which  was first 
 dubbed ``cosmic downsizing'' by Cowie et al. With the 
superb statistics provided by the SDSS we have been able to document the consequences 
of downsizing at the present epoch. 

It is particularly intriguing that the majority of black hole growth is occurring in 
galaxies lying just above the values of galaxy mass, density, and concentration where  
the galaxy population abruptly transitions from low density, disk-dominated
systems with ample ongoing star formation to dense, bulge-dominated 
systems with little on-going star-formation. This may imply that AGN today occur in a   
in a narrow ``habitable zone'' with a precipice to the low mass side (no black 
holes) and an on-coming famine at higher masses (no cold interstellar gas for 
fuel). 

Kauffmann et al. (2004) have speculated that the observed trends in galaxy
properties as a function of density can help us understand 
how galaxies evolve as a function of cosmic time.  According
to the standard cosmological paradigm , structure in the present-day
Universe formed through a process of hierarchical clustering, with
small structures merging to form progressively larger ones.  The
theory predicts that density fluctuations on galaxy scales collapsed
earlier in regions that are currently overdense.  Galaxies in high
density regions of the Universe such as galaxy clusters are thus more
`evolved' than galaxies in low density regions or voids.

Fig. 12 shows an example of a  rich system of galaxy groups at z=0.05 as surveyed by
the Sloan Digital Sky Survey. We have placed the 
origin of our x-and y--axes at the centre of the largest group in the field, which has a velocity
dispersion of $\sim 500$ km s$^{-1}$. We plot a  `slice' that includes
all galaxies with $cz$ within  500 km s$^{-1}$ of the brightest
galaxy in the group. The structures shown in Fig. 12  occupy a relatively
thin sheet that is aligned perpendicular to the line-of-sight.
As can be seen, the densest structures in the sheet appear to be arranged along
filaments that extend over scales of several tens of Mpc. 

The top left panel in Fig. 12 shows the distribution of all the galaxies in the slice  
with spectroscopic redshifts. In the top right panel we have colour-coded
galaxies with $10^{10} M_{\odot} < M_* < 3\times 10^{10} M_{\odot}$
according to their measured 4000 \AA\ break strengths. Red indicates galaxies with
D$_n$(4000)$> 1.8$, green is for galaxies with $1.6 <$ D$_n$(4000)$< 1.8$, and
blue is for galaxies with D$_n$(4000)$< 1.6$.
As can be seen, the mean stellar ages of galaxies are very strongly correlated
with local density.  
In the bottom panel we have colour-coded galaxies
with detected AGN according to their measured [OIII] line luminosities. The qualitative
picture is very similar.  AGN with high values of L[OIII] are found predominantly in low-density
regions, while low-luminosity AGN are also found in denser groups. Note that no AGN are found near
the very center of the most massive group in the field. 

These results again suggest that the cosmic evolution of star formation and of AGN activity
in the Universe are inextricably linked. What remains to be understood is whether
the AGN are  {\em responsible} for  active star formation
passing from massive galaxies at high redshifts to low  mass galaxies
at the present day, or whether  AGN have  simply responded to changes     
in their host galaxies and have no ultimate control over their evolution.       
  
\begin{figure}
\centerline{
\epsfxsize=14cm \epsfysize=14cm \epsfbox{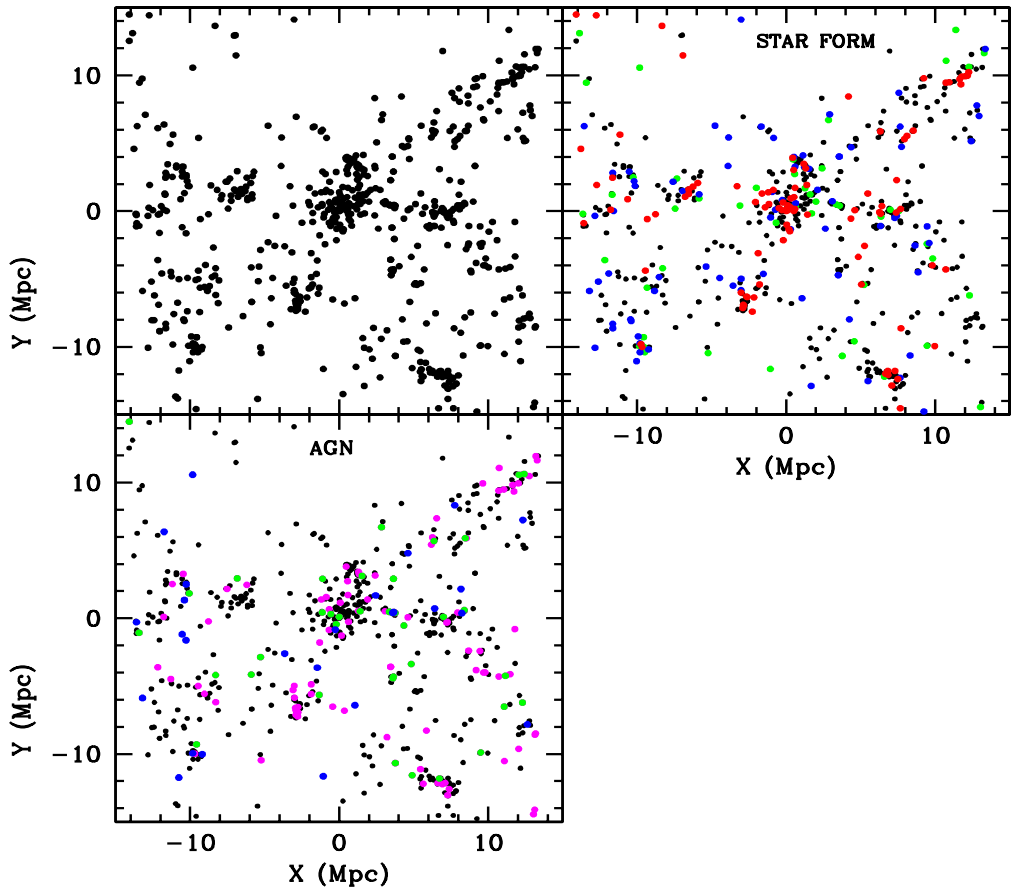}
}
\caption{\label{fig12}
\small
Top Left: The distribution of all galaxies in a `slice' at z=0.05 (see text for more details).
Top Right: Galaxies with $10^{10} M_{\odot} < M_* < 3\times 10^{10} M_{\odot}$ 
are colour-coded according
their measured 4000 \AA\ break strengths. Red is for D$_n$(4000)$>1.8$,
green is for $1.6 <$ D$_n$(4000)$<1.8$ and blue is for D$_n$(4000)$<1.6$.
Bottom: Galaxies hosting AGN are colour-coded according to L[OIII].
Magenta is for log L[OIII]$<6.5$, green is for $6.5<$log L[OIII] $<7$,
and blue is for log L[OIII]$>7$.}
\end {figure}
\normalsize

\vspace{1.0 cm}

Funding for the creation and distribution of the SDSS Archive has been provided
by the Alfred P. Sloan Foundation, the Participating Institutions, the National
Aeronautics and Space Administration, the National Science Foundation, the U.S.
Department of Energy, the Japanese Monbukagakusho, and the Max Planck Society.
The SDSS Web site is http://www.sdss.org/.

The SDSS is managed by the Astrophysical Research Consortium (ARC) for the
Participating Institutions. The Participating Institutions are The University
of Chicago, Fermilab, the Institute for Advanced Study, the Japan Participation
Group, The Johns Hopkins University, Los Alamos National Laboratory, the
Max-Planck-Institute for Astronomy (MPIA), the Max-Planck-Institute for
Astrophysics (MPA), New Mexico State University, University of Pittsburgh,
Princeton University, the United States Naval Observatory, and the University
of Washington.

\pagebreak 
\Large
\begin {center} {\bf References} \\
\end {center}
\normalsize
\parindent -7mm  
\parskip 3mm

Abazaijan, K. et al, 2003, AJ, 126, 2081

Antonucci, R. 1993, ARA\&A,31, 473

Baldwin, J., Phillips, M. \& Terlevich, R. 1981,
PASP, 93, 5

Balogh, M., Morris, S., Yee, H., Carlberg, R., \& Ellingson, E.
1999, ApJ, 527, 54

Blanton, M.R. et al, 2003a, ApJ, 594, 186 

Blanton, M.R., Lupton, R.H., Maley, F.M., 
Young, N., Zehavi, I., 
Loveday, J. 2003b, AJ, 125, 2276

Boyle, B.J., Shanks, T., Croom, S.M., Smith, R.J., 
Miller, L., Loaring, N. \& Heymans, C., 2000, MNRAS, 317, 1014
308, 77

Brinchmann, J., Charlot, S., White, S.D.M., Tremonti, C., Kauffmann, G., Heckman, T.M.,
Brinkmann, J., 2003, MNRAS, submitted (astro-ph/0311060)

Bruzual, G., Charlot, S., 2003, MNRAS, 344, 1000

Budavari, T., et al, 2003, ApJ, 595, 59 

Cattaneo, A., Haehnelt, M. \& Ress, M.J. 1999, MNRAS,
308, 77

Cid Fernandes, R. \& Terlevich, R. 1995, MNRAS, 272, 423

Cid Fernandes, R., Heckman, T., Schmitt, H., Gonz\'alez Delgado, R.,
\&  Storchi-Bergmann, T. 2001, ApJ, 558, 81

Cid Fernandes, R., Gonz\'alez-Delgado, R., Schmitt, H., Storchi-Bergmann, T., Martins, L.P.,
Perez, E., Heckman, T.M., Leitherer, C., Schaerer, D., 2004, ApJ, in press (astro-ph/0401416)

Colina, L., \& Arribas, S. 1999, ApJ, 514, 637 

Cowie, L.L., Songaila, A., Hu, E.M. and Cohem, J.G. 1996, AJ, 112, 839

Dahari,O. \& Robertis, M.M. 1988, ApJ, 331, 727

de Mello, D., Leitherer, C., \& Heckman, T. 2000, ApJ, 530, 251 

Dressler, A., \& Gunn, J. 1983, ApJ, 270, 7

Elvis, M. et al. 1994, ApJS , 95, 1

Fan, X. et al. 2001, AJ, 121, 31                            

Ferrarese, L. \& Merritt,D.  2000, ApJ, 539, L9     

Fukugita, M., Ichikawa, T., Gunn, J.E., Doi, M., Shimasaku, K., 
Schneider, D.P. 1996, AJ, 111, 1748

Gebhardt, K. et al 2000, ApJ , 543, L5     

Genzel, R., Pichon, C., Eckart, A., Gerhard, O.E.,
\& Ott, T. 2000, MNRAS, 317, 348

Gonz\'alez Delgado, R.,  Heckman, T., Leitherer, C., Meurer, G.,
Krolik, J., Wilson, A., Kinney, A., \& Koratkar, A. 1998, ApJ, 505, 174

Gonz\'alez Delgado, R., Heckman, T., \& Leitherer, C. 2001, ApJ, 546, 845

Gonz\'alez Delgado, R.,  Cid Fernandez, R., Perez, E., Martins, L.P., Storchi-Bergmann, T.,
Schmitt, H., Heckman, T.M., Leitherer, C., 2004,  ApJ, in press (astro-ph/0401414)

Granato, G.L., Silva, L., Monaco, P., Panuzzo, P.,
Salucci, P., De Zotti, G. \& Danese, L 2001, MNRAS, 324, 757

Gunn, J., Carr, M., Rockosi, C., Sekiguchi, M., Berry, K., Elms,
B., de Haas, E., Ivezi\'{c}, Z. et al. 1998, ApJ, 116, 3040

Heckman, T.M. 1980a, A\&A , 87, 142

Heckman, T.M. 1980b, A\&A , 87, 152

Heckman, T.M. et al. 1995, ApJ, 452, 549

Heckman, T.M., Gonz\'alez Delgado, R., Leitherer, C., Meurer, G., Krolik, J.,
Wilson, A., Koratkar, A., \& Kinney, A. 1997, ApJ, 482, 114

Heckman, T.M., Kauffmann, G., Brichmann, J., Charlot, S., Tremonti, C., White, S.D.M.,
2004, in preparation

Herrnstein,J.R., Moran, J.M.,Greenhill, L.J.,
Diamond, P.J., Inoue, M., Nakai, N., Miyoshi, M., Henkel, C., \& Riess, A. 1999,
Nature,400, 539

Hogg, D., Finkbeiner, D., Schlegel, D., \& Gunn, J. 2001, AJ, 122, 2129

Ho, L.C., Philippenko, A.V. \& Sargent, W.L.W. 1997, 
ApJS, 112, 315

Ho, L., Fillipenko, A., \& Sargent, W. 2003, ApJ, 583, 159                                         

Hubble, E.P., 1926, ApJ, 64, 321

Joguet, B., Kunth, D., Melnick, J., Terlevich, R., \& Terlevich, E. 2001,
A\&A, 380, 19

Koski, A. 1978, ApJ, 223, 56

Kauffmann, G. \& Haehenelt, M.G. 2000, MNRAS,
311,576

Kauffmann, G. et al, 2003a, MNRAS, 341, 33 (Paper I) 

Kauffmann, G. et al, 2003b, MNRAS, 341, 54 (Paper II) 

Kauffmann, G. et al 2003c, MNRAS, 346, 1055                         

Kauffmann, G. et al, 2004, MNRAS, submitted (astro-ph/0402030)

Kewley, L., Dopita, M., Sutherland, R., Heisler, C. \&
Trevena, J. 2001, ApJ, 556, 121

Leitherer, C. et al. 1999, ApJS, 123, 3

Lynden-Bell, D. 1969, MNRAS, 143, 167                

Madgwick, D.S. et al, 2003, MNRAS, 344, 847

Maoz, D., Koratkar, A., Shields, J., Ho, L., Filippenko, A., \& Sternberg, A.
1998, AJ, 116, 55

Marconi, A. \& Hunt, L.K., 2003, ApJ, 589, L21           

Marconi, A., Risaliti, G., Gilli, R.,
Hunt, L., Maiolino, R. \& Salvati, M. 2004, MNRAS, in press
(astro-ph/0311619

Nelson, C. \&  Whittle, M. 1999, AdSpR, 23, 891

Oliva, E., Origlia, L. Maiolino, R., \& Moorwood, A. 1999, A\&A , 350, 9

Pier, J.R., Munn, J.A., Hindsley, R.B., Hennessy, G.S., Kent, S.M.,
Lupton, R.H., Ivezi\'{c}, Z., 2003, AJ, 125, 1559

Richards, G. et al. 2002, AJ, 123, 2945

Salpeter, E. 1964, ApJ, 140, 796                       

Schinnerer, E., Colbert, E., Armus, L., Scoville, N., \& Heckman, T. 2003,
Coevolution 
of Black Holes and Galaxies, ed. L. C. Ho (Pasadena: Carnegie Observatories,
http://www.ociw.edu/ociw/symposia/series/symposium1/proceedings.html

Schmitt, H., Storchi-Bergmann, T., \& Cid Fernandes, R. 1999, MNRAS, 303, 173

Smith, J.A., et al 2002, AJ, 123, 2121

Shimasaku, K. et al, 2001, AJ, 122, 1238                                                

Steffen, A.T., Barger, A.J., Cowie, L.L.,
Mushotsky, R.F. \& Yang, Y. 2003, ApJ, 596, L23

Stoughton, C. et al, 2002, AJ, 123, 485

Strateva, I. et al, 2001, AJ, 122, 1861                                               

Strauss, M., et al 2002, AJ, 124, 1810  

Terlevich, E., D\'iaz, A., \& Terlevich, R. 1990, MNRAS, 242, 271

Tran, H. 1995, ApJ, 440, 597

Tremaine, S. et al. 2002, ApJ, 574, 740

Ueda,Y., Masayuki, A., Ohta, K. \& Miyaji, T. 2003,
ApJ, 598, 886

Veilleux, S. \& Osterbrock, D. 1987,
ApJS, 63, 295

Worthey, G. \& Ottaviani, D.L., 1997, ApJS, 111,377

York D.G. et al, 2000, AJ, 120, 1579                                                             

Yu, Q. \& Tremaine, S. 2002, ApJ, 335, 965             

Zakamska, N. et al. 2003, AJ, 126, 2125

Zehavi, I. et al., 2002, ApJ, 571, 172

\end {document}